\documentclass[prl,twocolumn,superscriptaddress,showpacs,amsmath,amssymb,floatfix]{revtex4}
\usepackage{graphicx,color}
\usepackage{epsfig}
\usepackage{bm}

\begin{document}

\title{Thermodynamic Bounds on  Efficiency
\\for Systems with Broken Time-reversal Symmetry}

\author{Giuliano Benenti}
\affiliation{CNISM, CNR-INFM \& Center for Nonlinear and Complex Systems,
Universit\`a degli Studi dell'Insubria, Via Valleggio 11, 22100 Como, Italy}
\affiliation{Istituto Nazionale di Fisica Nucleare, Sezione di Milano,
via Celoria 16, 20133 Milano, Italy}
\author{Keiji Saito}
\affiliation{Department of Physics, Graduate School of Science,
University of Tokyo, Tokyo 113-0033, Japan}
\author{Giulio Casati}
\affiliation{CNISM, CNR-INFM \& Center for Nonlinear and Complex Systems,
Universit\`a degli Studi dell'Insubria, Via Valleggio 11, 22100 Como, Italy}
\affiliation{Istituto Nazionale di Fisica Nucleare, Sezione di Milano,
via Celoria 16, 20133 Milano, Italy}
\affiliation{Centre for Quantum Technologies,
National University of Singapore, Singapore 117543}
\date{\today}


\begin{abstract}
We show that for systems with broken time-reversal symmetry  the
maximum efficiency and  the efficiency at maximum power are both determined 
by two parameters: a ``figure of merit'' and an asymmetry parameter.
In contrast to the time-symmetric case, the figure of merit is
bounded from above; nevertheless the Carnot efficiency can be reached 
at lower and lower values of the figure of merit and far from the
so-called strong coupling condition as the asymmetry
parameter increases.
Moreover, the Curzon-Ahlborn limit for efficiency at maximum
power can be overcome within linear response. Finally, always within linear response, it is allowed to have simultaneously Carnot efficiency and non-zero power.
\end{abstract}
\maketitle

The understanding of the fundamental limits that thermodynamics imposes on the efficiency of thermal machines is a central issue in physics and is becoming more and more practically relevant in the future society. In particular due to the need of providing a sustainable supply of energy
and to strong concerns about the environmental
impact of the combustion of fossil fuels, there is an increasing pressure
to find best thermoelectric
materials~\cite{mahanPT,majumdar,dresselhaus,snyder}.

A cornerstone result goes back to Carnot~\cite{carnot}. In a cycle between two reservoirs at temperatures $T_1$ and $T_2$ $(T_1>T_2)$, the efficiency $\eta$,  defined as the ratio  of the performed work $W$ over the heat $Q_1$ extracted from the high temperature reservoir, is bounded by the so-called Carnot efficiency $\eta_C$:
\begin{equation}
\eta = W/Q_1 \leq \eta_C = 1-T_2/T_1.
\end{equation}
The Carnot efficiency is obtained for a quasi static transformation which requires infinite time and therefore the extracted power, in this limit, reduces to zero. For this reason the notion of efficiency at maximum power has been introduced.

An upper bound for the efficiency at maximum power has been proposed long ago by several authors
~\cite{Yvon,chambadal,novikov,curzon} and is commonly referred to 
as Curzon-Ahlborn upper bound:
\begin{equation}
\eta_{CA} = 1-\sqrt{T_2/T_1}.
\end{equation}
The range of validity of this bound has been widely discussed in several interesting papers~\cite{vandenbroeck,espositoa,schulman,esposito,linke,seifert}.
For the thermoelectric power generation and refrigeration, within linear response
and for systems with time-reversal symmetry,
both the maximum efficiency and the efficiency at maximum power, are governed by a single parameter, the
dimensionless figure of merit
\begin{equation}
ZT=\frac{\sigma S^2}{\kappa}\, T,
\end{equation}
where $\sigma$ is the electric conductivity, $S$ is the
thermoelectric power (Seebeck coefficient),
$\kappa$ is the thermal conductivity, and $T$ is the
temperature.
The maximum efficiency is given by
\begin{equation}
\eta_{\rm max}=
\eta_C\,
\frac{\sqrt{ZT+1}-1}{\sqrt{ZT+1}+1},
\label{etamaxB0}
\end{equation}
where $\eta_C$ is the Carnot efficiency;
the efficiency $\eta(\omega_{\rm max})$ at maximum power $\omega_{\rm max}$ reads~\cite{vandenbroeck}
\begin{equation}
\eta(\omega_{\rm max})=\eta_{CA}^{(l)}\,\frac{ZT}{ZT+2}.
\label{etawmaxB0}
\end{equation}
The only restriction imposed by thermodynamics is $ZT\ge 0$, so that
$\eta_{\rm max}\le \eta_C$ and $\eta (\omega_{\rm max})\le 
\eta_{CA}^{(l)}$,
where $\eta_{CA}^{(l)}=\eta_C/2$ is the Curzon-Alhborn efficiency
in the linear response regime.
The upper bounds $\eta_C$ and $\eta_{CA}^{(l)}$ are reached when 
the figure of merit $ZT\to\infty$.
This limit corresponds to the so-called strong coupling condition,
for which the Onsager matrix ${\bm L}$ 
becomes singular (that is, $\det {\bm L}=0$) and therefore 
the ratio $J_q/J_\rho$, with $J_q$ heat currrent and $J_\rho$ 
electric (particle) current, is independent of the 
applied temperature and chemical potential gradients.

In this Letter we investigate, within the linear response regime,
the case when time-reversal symmetry
is broken, for instance by means of an applied magnetic field~\cite{harman}. We show
that in this case  the maximum
efficiency  as well as the
efficiency at maximum power depend on two parameters: the first parameter is a
generalization of the figure of merit $ZT$, while the second,
asymmetry parameter, is the
ratio of the off-diagonal
terms of the Onsager matrix. The presence of a second parameter is highly important since it offers
an additional freedom in the design of high-performance thermoelectric devices.
In particular it turns out that the figure of merit is bounded from above 
when the asymmetry parameter is different from unity; 
nevertheless the Carnot efficiency is reached at lower and lower 
values of the figure of merit and far from the strong coupling 
condition as the asymmetry parameter increases.
Furthermore, the Curzon-Ahlborn limit can be overcome.
Finally, within linear response it is not forbidden
to have simultaneously Carnot efficiency and non-zero power.

\begin{figure}
\begin{center}
\epsfxsize=80mm\epsffile{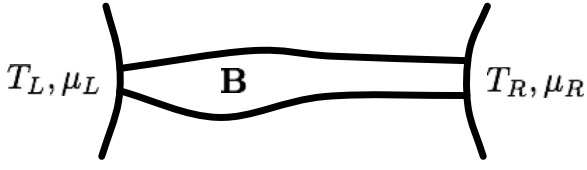}
\caption{Schematic drawing of the model.}
\label{fig:scheme}
\end{center}
\end{figure}

The model we consider is sketched in Fig.~\ref{fig:scheme}.
Both electric and heat currents flow along the horizontal axis. The system is in contact with left and right
reservoirs at temperatures $T_L$ and $T_R$ and chemical potentials
$\mu_L$ and $\mu_R$.
Even though fluxes are one-dimensional, the motion
inside the system can be two- or three-dimensional.
We start from the equations connecting fluxes and thermodynamic forces
within linear irreversible thermodynamics~\cite{callen}:
\begin{equation}
\left\{
\begin{array}{l}
{\displaystyle
J_\rho({\bm B})=L_{\rho\rho} ({\bm B}) X_1 + L_{\rho q}({\bm B}) X_2,
}
\\
\\
{\displaystyle
J_q({\bm B})=L_{q\rho} ({\bm B}) X_1 + L_{q q}({\bm B}) X_2,
}
\end{array}
\right.
\end{equation}
where $J_\rho$ and $J_q$ are the particle and heat currents,
${\bm B}$ an applied magnetic field or any parameter breaking
time-reversibility (such as the Coriolis force, etc.), and
$X_1=-\beta\Delta\mu$, $X_2=\Delta\beta=-\Delta T/T^2$
the thermodynamic forces, with
$\Delta\mu=\mu_R-\mu_L$, $\beta=1/T$,
$\Delta\beta=\beta_R-\beta_L$.
$\Delta T=T_R-T_L$ is assumed to be small compared to 
$T_L\approx T_R \approx T$.
Without loss of generality we assume
$T_L>T_R$. Therefore, the parameter $X_2$ is
always positive, while the sign of $X_1$ is
determined in such a way that the work done by the
particle current is positive.
Note that the sign of the current is taken positive
if it flows from the left to the right reservoir.

The positivity of the entropy production rate,
\begin{equation}
\dot{S}=J_\rho X_1 + J_q X_2\ge 0,
\end{equation}
implies for the Onsager coefficients $L_{ij}$ ($i,j=\rho,q$) that
\begin{equation}
\left\{
\begin{array}{l}
{\displaystyle
L_{\rho\rho}({\bm B})\ge 0,
}
\\
{\displaystyle
L_{qq}({\bm B})\ge 0,
}
\\
{\displaystyle
L_{\rho\rho}({\bm B})L_{qq}({\bm B})
-\frac{1}{4}[L_{\rho q}({\bm B})+L_{q \rho}({\bm B})]^2 \ge 0.
}
\end{array}
\right.
\label{dots}
\end{equation}
Moreover, the Onsager-Casimir relations in the presence of a magnetic field read
\begin{equation}
L_{ij}({\bm B})=L_{ji}(-{\bm B}).
\end{equation}

The Onsager coefficients are related to the familiar transport coefficients
$\sigma$, $\kappa$, $S$ as follows~\cite{callen}:
\begin{equation}
\sigma({\bm B})=\frac{e^2}{T}\,L_{\rho\rho}({\bm B}),
\end{equation}
\begin{equation}
\kappa({\bm B})=\frac{1}{T^2}\frac{\det {\bm L}({\bm B})}{L_{\rho\rho}({\bm B})},
\end{equation}
\begin{equation}
S({\bm B})=\frac{L_{\rho q}({\bm B})}{e T L_{\rho\rho}({\bm B})},
\;\;\;\;\;
S(-{\bm B})=\frac{L_{q\rho}({\bm B})}{e T L_{\rho\rho}({\bm B})},
\end{equation}
where $e$ is the electron charge and ${\bm L}$ denotes
the Onsager matrix with matrix elements
$L_{ij}$.
Note that the Onsager-Casimir relations
$L_{ij}({\bm B})=L_{ji}(-{\bm B})$ imply $\sigma({\bm B})=\sigma(-{\bm B})$
and $\kappa({\bm B})=\kappa(-{\bm B})$, while a priori it is possible to
have $S({\bm B})\ne S(-{\bm B})$.
In what follows, to improve readibility we do not write 
${\bm B}$ explicitly as argument in the
Onsager coefficients, unless necessary.

\textit{Efficiency at maximum power.}
The efficiency $\eta$, under steady-state conditions, is given by the ratio
of the output power over the heat current (leaving the hot reservoir):
\begin{equation}
\eta=\frac{\omega}{J_q}.
\label{eq:etadef}
\end{equation}
The output power
\begin{equation}
\omega = J_\rho \Delta \mu = - J_\rho T X_1
\end{equation}
is maximal when
\begin{equation}
X_1 = -\frac{L_{\rho q}}{2 L_{\rho\rho}} X_2
\end{equation}
and is given by
\begin{equation}
\omega_{\rm max} = \frac{T}{4} \frac{L_{\rho q}^2}{L_{\rho\rho}} X_2^2
=\frac{\eta_C}{4} \frac{L_{\rho q}^2}{L_{\rho\rho}} X_2,
\label{eq:omegamax}
\end{equation}
where $\eta_C=-\Delta T/T=TX_2$ is the Carnot efficiency.

The efficiency at maximum power
\begin{equation}
{\displaystyle
\eta(\omega_{\rm max}) = \frac{\omega_{\rm max}}{J_q}
= \eta_{CA}^{(l)}\, \frac{1}{2\frac{L_{\rho\rho}L_{qq}}{L_{\rho q}^2}
-\frac{L_{q\rho}}{L_{\rho q}}}.
}
\end{equation}

\begin{figure}
\begin{center}
\epsfxsize=80mm\epsffile{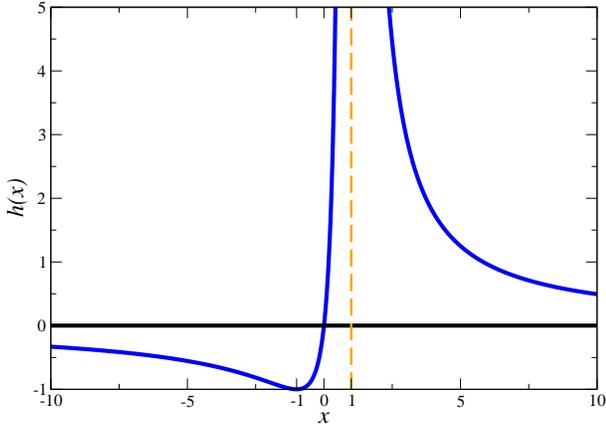}
\caption{Function $h(x)$ (blue solid curve).
This function has a vertical asymptote at $x=1$
(dashed line).
Thermodynamics restricts the parameter
$y$ between $y=0$ and $y=h$.}
\label{fig:hx}
\end{center}
\end{figure}
\begin{figure}
\begin{center}
\epsfxsize=80mm\epsffile{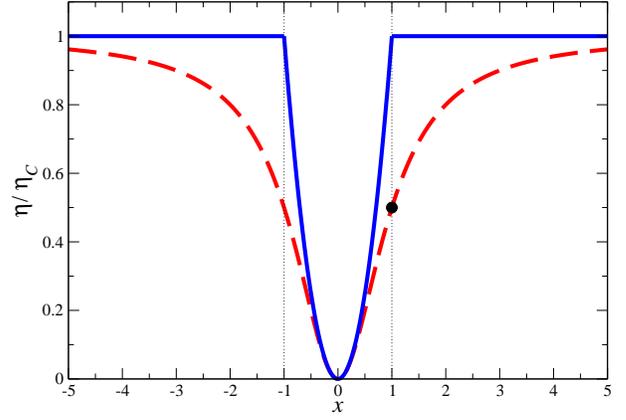}
\caption{Ratio $\eta/\eta_C$ as a function of the asymmetry parameter $x$,
with $\eta=\eta^\star$ (dashed curve) and $\eta=\eta_M$
(full curve). The black circle corresponds to the Curzon-Ahlbohrn
limit at $x=1$: $\eta_{CA}^{(l)}=\eta_C/2$.}
\label{fig:eta}
\end{center}
\end{figure}

is seen to depend on two parameters:
\begin{equation}
x\equiv\frac{L_{\rho q}}{L_{q \rho}}=\frac{S({\bm B})}{S(-{\bm B})},
\end{equation}
\begin{equation}
y\equiv\frac{L_{\rho q}L_{q \rho}}{\det {\bm L}}=
\frac{\sigma({\bm B}) S({\bm B})S(-{\bm B})}{\kappa({\bm B})}\,T.
\end{equation}
and writes
\begin{equation}
\eta(\omega_{\rm max})=
\eta_{CA}^{(l)}\,\frac{xy}{2+y}.
\label{etawmax}
\end{equation}
In the particular case $x=1$, $y$ reduces to the $ZT=(\sigma S^2/k)T$
figure of merit of the
time-symmetric case and Eq.~(\ref{etawmax}) reduces to
Eq.~(\ref{etawmaxB0}).
While thermodynamics does not impose any restriction on the attainable
values of the asymmetry parameter $x$,
the third inequality in (\ref{dots}) implies
\begin{equation}
\left\{
\begin{array}{l}
h(x)\le y \le 0 \;\; {\rm if}\,\, x<0,
\\
\\
0\le y \le h(x) \;\; {\rm if}\,\, x>0,
\end{array}
\right.
\label{ybounds}
\end{equation}
where $h(x)=4x/(x-1)^2$ and we have taken into account that
$x$ and $y$ must have the same sign since (\ref{dots})
implies $\det {\bm L}\ge 0$
and $y=xL_{q\rho}^2/\det{\bm L}$.
The function $h(x)$ is drawn in Fig.~\ref{fig:hx}.
Note that $\lim_{x\to 1}h(x)=\infty$ and
therefore there is no upper bound on
$y({x=1})=ZT$.
It is easy to check that
the maximum $\eta^\star$ in (\ref{etawmax}) is achieved for $y=h(x)$, that is,
\begin{equation}
\eta(\omega_{\rm max})\le \eta^\star= \eta_C\,
\frac{x^2}{x^2+1}.
\label{eq:etastar}
\end{equation}
The function $\eta^\star (x)$ is drawn in Fig.~\ref{fig:eta} (dashed curve).
Several remarks are in order. To begin with, in the absence of the magnetic
field ($x=1$) $L_{\rho q}=L_{q \rho}$  and the Curzon-Ahlborn limit
for the linear response regime is recovered:
$\eta^\star(x=1)=\eta_{CA}^{(l)}=\eta_C/2$.
Furthermore, the Curzon-Ahlborn limit can be overcome when $|x|>1$
and $\eta^\star$ approaches the Carnot efficiency when
$|x|\to\infty$.
We also note that if the magnetic field ${\bm B}$ is reversed,
owing to the Onsager-Casimir relations,
$x$ is replaced by $1/x$. From inequality (\ref{eq:etastar}) it then follows that
the average efficiency
for ${\bm B}$ and $-{\bm B}$ cannot overcome the
Curzon-Ahlborn limit: $\frac{1}{2}(\eta^\star(x)+\eta^\star(1/x))\le
\eta_{CA}^{(l)}$.

\textit{Maximum efficiency.}
The maximum of
\begin{equation}
\eta=\frac{\Delta \mu J_\rho}{J_q}=
\frac{-TX_1(L_{\rho\rho}X_1+L_{\rho q} X_2)}{L_{q\rho}X_1+L_{qq}X_2}
\end{equation}
over $X_1$, for fixed $X_2$ and under the condition
$J_q>0$, is achieved for
\begin{equation}
X_1=\frac{L_{qq}}{L_{q\rho}}
\left(-1+\sqrt{\frac{\det {\bm L}}{L_{\rho\rho} L_{qq}}}
\right) X_2
\end{equation}
and is given by
\begin{equation}
\eta_{\rm max}= \eta_C\,x\,
\frac{\sqrt{y+1}-1}{\sqrt{y+1}+1}.
\label{eq:ZTx}
\end{equation}
Note that (\ref{ybounds}) implies $y\ge -1$ for any $x$, so that
$\eta_{\rm max}$ is as expected a real-valued function.
We point out that at $x=1$ we recover the well-known efficiency
expression (\ref{etamaxB0}).
For a given asymmetry parameter $x$ the maximum $\eta_M$ of (\ref{eq:ZTx})
is again reached when $y=h(x)$. By substituting the function $h(x)$
into Eq.~(\ref{eq:ZTx}) we find
\begin{equation}
\eta_M=
\left\{
\begin{array}{ll}
\eta_C\,x^2 & {\rm if}\,\, |x| \le 1,
\\
\\
\eta_C & {\rm if}\,\, |x| \ge 1.
\end{array}
\right.
\end{equation}
The function $\eta_M(x)$ is drawn in Fig.~\ref{fig:eta} (full curve).
On the other hand, when $x\ne 1$ the figure of merit $y$ alone is no longer
sufficient to determine the thermoelectric efficiency:
$\eta_{\rm max}$ depends on both $x$ and $y$.
Moreover, the Carnot limit can be achieved only when
$|x|\ge 1$~\cite{footnote}. We point out that when $|x|\to\infty$, the figure
of merit $y$ required to get the Carnot efficiency becomes
increasingly smaller.
When $|x|\ge 1$ the Carnot efficiency is obtained under the condition
$y=h(x)$, which implies $\det {\bm L}=(L_{\rho q}-L_{q\rho})^2/4$.
Therefore Carnot efficiency and $L_{\rho q}\ne L_{q \rho}$ imply
$\det {\bm L}>0$, that is, the strong coupling condition
is not fulfilled.

The entropy production rate at maximum efficiency is
\begin{equation}
\dot{S}(\eta_M)=
\left\{
\begin{array}{ll}
{\displaystyle
\frac{(L_{\rho q}^2-L_{q\rho}^2)^2}{4L_{\rho\rho}L_{q\rho}^2}\, X_2^2} & {\rm if} \,\, |x|< 1,
\\
\\
0 & {\rm if} \,\, |x|\ge 1.
\end{array}
\right.
\label{dotS}
\end{equation}
Hence there is no entropy production at $|x|\ge 1$, in agreement
with the fact that in this regime $\eta_M=\eta_C$.

We can now derive the output power at maximum efficiency:
\begin{equation}
\omega(\eta_M)=\frac{\eta_M}{4}\,
\frac{|L_{\rho q}^2 - L_{q\rho}^2|}{L_{\rho\rho}}\, X_2.
\label{eq:omegaM}
\end{equation}
From relation (\ref{eq:etadef}), the heat current is determined as
$J_q = |L_{\rho q}^2 - L_{q \rho}^2 |
X_2 / (4 L_{\rho \rho}) $.
It is readily seen from (\ref{eq:omegamax}) and (\ref{eq:omegaM})
that $\omega(\eta_M)\le \omega_{\rm max}$. It is important to note
that $\omega(\eta_M)\to \omega_{\rm max}$ when $|x|\to\infty$,
as expected since in this limit $\eta^\star\to \eta_M=\eta_C$.
Therefore, in this limit we have  Carnot efficiency and 
power $\omega_{\rm max}$ simultaneously.

In summary, we have shown that when time-reversal symmetry is broken
both the maximum efficiency and the
efficiency at maximum power are no longer exclusively determined by
the figure of merit $ZT$. Two parameters are needed, an asymmetry
parameter $x$ and a parameter $y$ which reduces to $ZT$ in the
symmetric limit $x=1$. In the case $|x|>1$, it is possible to overcome
the Curzon-Ahlborn limit within linear response and to reach the
Carnot efficiency, for increasingly smaller and smaller figure of merit $y$ as
$|x|$ becomes larger. With regard to the practical relevance of the
results presented here, we should note that in the non-interacting case 
$S({\bm B})=S(-{\bm B})$, thus implying $x=1$, is a consequence
of the symmetry properties of the scattering matrix~\cite{datta}.
On the other hand, the Onsager-Casimir symmetry relations do not impose 
the symmetry of the Seebeck coefficient under the exchange
${\bm B}\to -{\bm B}$.
Therefore, this symmetry 
may be violated when electron-phonon and electron-electron
interactions are taken into account.
While the Seebeck coefficient has always been found 
to be an even function of the magnetic
field in two-terminal purely metallic mesoscopic 
systems~\cite{vanlangen}, 
Andreev interferometer experiments~\cite{chandrasekhar}
and recent theoretical studies~\cite{jacquod,imry} 
have shown that systems
in contact with a superconductor or with a
heat bath can exhibit non-symmetric thermopowers.
It is a challenging problem to find realistic setups with $x$ 
significantly different from unity, while approaching the Carnot
efficiency.

Useful discussions with Yoseph Imry are gratefully
acknowledged.
This work has been initiated during the Advanced Study Group on Thermodynamics of Finite Systems
at the Max Planck Institute for the Physics of Complex Systems, Dresden.
We acknowledge the support by the
MIUR-PRIN 2008 and by Regione Lombardia;
KS was supported by MEXT, Grant Number (21740288).


\begin{thebibliography}{0}
\bibitem{footnote}
It is interesting the fact that,
in the case of a refrigerator, the maximum efficiency
$\eta^{(r)}=J_q/\omega$ ($J_q<0$, $\omega<0$),
achieved for
\begin{equation}
X_1=\frac{L_{qq}}{L_{q\rho}}
\left(-1-\sqrt{\frac{\det {\bm L}}{L_{\rho\rho} L_{qq}}}
\right) X_2,
\end{equation}
is given by
\begin{equation}
\eta_{\rm max}^{(r)}= \eta_C\,\frac{1}{x}\,
\frac{\sqrt{y+1}-1}{\sqrt{y+1}+1}.
\end{equation}
\end{thebibliography}
\end{document}